\documentclass[runningheads]{llncs}

\usepackage[T1]{fontenc}

\usepackage{graphicx}
\usepackage{siunitx}
\usepackage{xcolor}
\usepackage{hyperref}
\usepackage{booktabs}
\newcommand*\samethanks[1][\value{footnote}]{\footnotemark[#1]}
\usepackage{color}

\newcommand{\bftab}{\fontseries{b}\selectfont}
\begin{document}
\title{Denoising Diffusion Models for 3D Healthy Brain Tissue Inpainting}

\author{Alicia Durrer$^{1}$ \and Julia Wolleb$^{1}$ \and Florentin Bieder$^{1}$ \and Paul Friedrich$^{1}$ \and Lester Melie-Garcia$^{1,2}$ \and Mario Ocampo-Pineda$^{1,2}$ \and Cosmin I. Bercea$^{3,4}$ \and Ibrahim E. Hamamci$^{5}$ \and Benedikt Wiestler$^{6}$ \and Marie Piraud$^{7}$  \and {\"O}zg{\"u}r Yaldizli$^{1,2}$ \and Cristina Granziera$^{1,2}$ \and Bjoern H. Menze$^{5}$ \and Philippe C. Cattin$^{1}$\thanks{equal contribution} \and Florian Kofler$^{6,7,8,9}$\samethanks
}

\authorrunning{Durrer et al.}

\institute{$^{1}$ Department of Biomedical Engineering, University of Basel, Switzerland\\
$^{2}$ University Hospital Basel, Switzerland\\
$^{3}$ Computational Imaging and AI in Medicine, Technical University of Munich, Germany\\
$^{4}$ Institute of Machine Learning in Biomedical Imaging, Helmholtz Center Munich, Germany\\
$^{5}$ Department of Quantitative Biomedicine, University of Zurich, Switzerland\\
$^{6}$ Department of Diagnostic and Interventional Neuroradiology, School of Medicine, Klinikum rechts der Isar, Technical University of Munich, Germany\\
$^{7}$ Helmholtz AI, Helmholtz Munich, Germany\\
$^{8}$ Department of Computer Science, TUM School of Computation, Information and Technology, Technical University of Munich, Germany\\
$^{9}$ TranslaTUM - Central Institute for Translational Cancer Research, Technical University of Munich, Germany\\
\email{alicia.durrer@unibas.ch}}

\maketitle              

\begin{abstract}
Monitoring diseases that affect the brain's structural integrity requires automated analysis of magnetic resonance (MR) images, e.g., for the evaluation of volumetric changes. 
However, many of the evaluation tools are optimized for analyzing healthy tissue. To enable the evaluation of scans containing pathological tissue, it is therefore required to restore healthy tissue in the pathological areas. In this work, we explore and extend denoising diffusion models for consistent inpainting of healthy 3D brain tissue. We modify state-of-the-art 2D, pseudo-3D, and 3D methods working in the image space, as well as 3D latent and 3D wavelet diffusion models, and train them to synthesize healthy brain tissue. Our evaluation shows that the pseudo-3D model performs best regarding the structural-similarity index, peak signal-to-noise ratio, and mean squared error. To emphasize the clinical relevance, we fine-tune this model on data containing synthetic MS lesions and evaluate it on a downstream brain tissue segmentation task, whereby it outperforms the established \textit{FMRIB Software Library (FSL) lesion-filling} method.

\end{abstract}

\section{Introduction}
\begin{figure}
\centering
\includegraphics[width=0.85\textwidth]{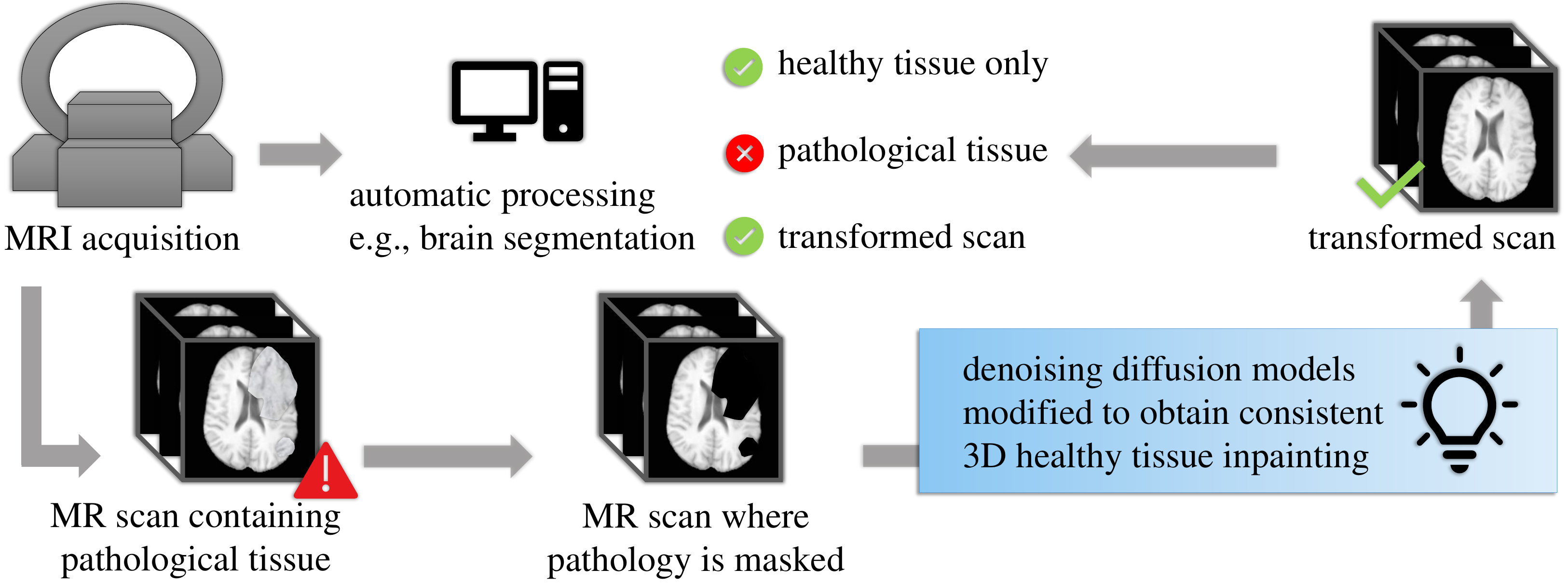}
\caption{Automatic MR image processing tools are often designed to evaluate healthy tissue only. Therefore, pathological tissue needs to be replaced by healthy tissue. We modify and evaluate 2D, pseudo-3D and 3D denoising diffusion models to obtain consistent 3D healthy tissue inpainting.}\label{overview_problem}
\end{figure}

Magnetic Resonance (MR) imaging is a valuable tool for disease monitoring. However, many automatic MR image evaluation tools are optimized for analyzing healthy tissue only  \cite{seiger2018cortical}. Pathological structures, such as brain tumors or Multiple Sclerosis (MS) lesions, impair this analysis and have to be masked out and replaced with healthy tissue during preprocessing \cite{almansour2021high,battaglini2012evaluating,dadar2021beware}.
Applying Denoising Diffusion Probabilistic Models (DDPMs) \cite{ho2020denoising,nichol2021improved} for healthy brain tissue inpainting \cite{durrer2024denoising} showed promising results in the "BraTS 2023 Inpainting Challenge"  \cite{kofler2023brain}.
However, the presented 2D slice-wise method \cite{durrer2024denoising} requires improvement as it suffers from stripe artifacts due to the lack of consistency between neighboring 2D slices.
Solving the inpainting task directly with 3D diffusion models is challenging due to their substantial memory requirements and limited scalability when dealing with high-resolution inputs. In this work, we extend and evaluate different state-of-the-art diffusion models for a 3D healthy brain tissue inpainting task and provide a comprehensive overview of the evaluated methods. Fig.~\ref{overview_problem} illustrates the task; the code is available at \url{https://github.com/AliciaDurrer/DM_Inpainting}.

\subsection{Related Work}
Throughout the last years, DDPMs have demonstrated state-of-the-art performance for image generation. 
They form the basis of well-known text-to-image generation frameworks, like "Stable Diffusion" \cite{rombach2022high}, and have successfully been applied to image-to-image translation tasks \cite{saharia2022palette}.
DDPMs outperform generative adversarial nets \cite{goodfellow2014generative} and methods based on variational autoencoders (VAE) \cite{razavi2019generating,van2017neural} for image synthesis tasks \cite{dhariwal2021diffusion}. Recent works, such as \cite{lugmayr2022repaint,saharia2022palette}, showed that they also manage to obtain better results than GAN-based \cite{suvorov2022resolution,yu2018generative,zeng2022aggregated} and VAE-based methods \cite{peng2021generating} developed for 2D inpainting. However, consistent 3D inpainting remains an unsolved challenge \cite{durrer2024denoising}. Nevertheless, diffusion models have also proven to be successful in the medical context, e.g., for segmentation \cite{wolleb2022segmentation}, implant generation \cite{friedrich2023point}, contrast harmonization \cite{durrer2023diffusion} and anomaly detection \cite{wolleb2022diffusion}. Due to their high memory requirements, they possess limited scalability from 2D to 3D. However, strategies to mitigate this issue have emerged, including the development of memory-efficient architectures in the image domain \cite{bieder2023memory}, as well as in the form of latent diffusion models (LDMs) \cite{khader2022medical,rombach2022high}. LDMs apply the diffusion model on a learned low-dimensional representation of the input data and thereby allow to reduce the sampling time of DDPMs. Sampling efficiency can also be enhanced by employing alternative sampling strategies, as elucidated in \cite{karras2022elucidating}, facilitating faster sampling rates. A recent approach, allowing to reduce both memory consumption and sampling time, is using generative models on the wavelet coefficients of the input images \cite{phung2023wavelet}. \cite{friedrich2024wdm} obtained promising results using 3D wavelet diffusion models for high-resolution medical image synthesis. 

\subsection{Contribution}
Existing diffusion models for inpainting create convincing 2D results, but their scalability to 3D remains a challenge. Therefore, we explore and modify state-of-the-art methods for image synthesis or image-to-image translation to solve a 3D inpainting task. Our work provides a comprehensive overview of the 2D, pseudo-3D, and 3D methods modified for 3D inpainting of healthy brain tissue. We train the different methods on the publicly available BraTS 2023 dataset. We find that the pseudo-3D model produces the qualitatively and quantitatively best results in terms of structural similarity index measure (SSIM), mean squared error (MSE), and peak signal-to-noise ratio (PSNR).
To demonstrate the clinical relevance of an effective healthy tissue inpainting method, we additionally evaluate the performance of the pseudo-3D model on a downstream segmentation task using a prospective longitudinal MS database with MR images showing MS lesions. The proposed pseudo-3D method outperforms the established \textit{FMRIB Software Library (FSL)} lesion-filling method \cite{battaglini2012evaluating}.

\section{Methods}
\subsection{Denoising Diffusion Probabilistic Models}
DDPMs are generative models that consist of an iterative forward process $q$ and a learned reverse process $p_{\theta}$. The forward process $q$ adds noise to an input image $x_0$ for $T$ time steps $t$. It can be formulated as a Markov chain, with each transition being a Gaussian  
\begin{equation}
q(x_{t}|x_{t-1}):=\mathcal{N}(x_{t};\sqrt{1-\beta _{t}}x_{t-1},\beta _{t}\mathbf{I}),
\label{eqn1}
\end{equation}
with the identity matrix $\mathbf{I}$ and the fixed forward process variances $\beta_{1},\ldots,\beta_{T}$.
The reverse process $p_{\theta}$ follows a sequence of Gaussian distributions
\begin{equation}
p_{\theta}(x_{t-1}\vert x_t):= \mathcal{N}\bigl(x_{t-1};\mu_{\theta}(x_t, t), \Sigma_\theta(x_t,t)\bigr),
\label{eqn2}
\end{equation}
with mean $\mu_{\theta}$ and variance $\Sigma_\theta$, parameterized by a time-conditioned model $\epsilon_{\theta }(x_{t},t)$. We train $\epsilon_{\theta }(x_{t},t)$ to predict the noise $\epsilon$ to be removed from a noisy image $x_t$ obtained from
\begin{equation}
x_{t}=\sqrt[]{\overline{\alpha} _{t}}x_{0}+\sqrt[]{1-\overline{\alpha} _{t}}\epsilon, \quad \mbox{with } \epsilon \sim \mathcal{N}(0,\mathbf{I}),
\label{eqn3}
\end{equation}
with $\alpha _{t}:=1-\beta _{t}$, $\overline{\alpha}_{t}:=\prod_{s=1}^t \alpha _{s}$. The training objective is given by the MSE loss between $\epsilon$ and $\epsilon_{\theta }(x_{t},t)$. Sampling can be done using the trained model $\epsilon_{\theta }$ in 
\begin{equation}
x_{t-1}=\frac{1}{\sqrt[]{\alpha _{t}}}\left(x_{t}-\frac{1-\alpha _{t}}{\sqrt[]{1-\overline{\alpha }_{t}}}\epsilon _{\theta }(x_{t},t)\right)+\sigma_{t}\mathbf{z}, \quad \mbox{with } \mathbf{z} \sim \mathcal{N}(0,\mathbf{I}),
\label{eqn4}
\end{equation}
starting from $x_T$  $\sim \mathcal{N}(0,\mathbf{I})$ and applying Eq.~\ref{eqn4} for all $t = T,\ldots,1$.

\subsection{Modifying Diffusion Models for Inpainting}
\begin{figure}
\centering
\includegraphics[width=0.93\textwidth]{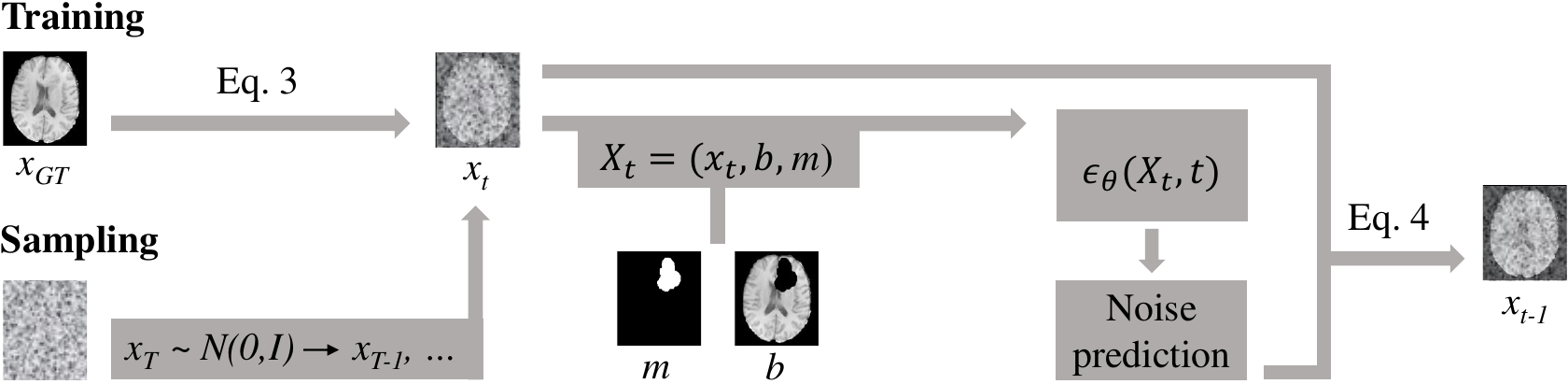}
\caption{We present an overview of the denoising process adapted for the inpainting task. For conditioning, the mask $m$ and the masked image $b$ are concatenated to the noisy image $x_t$, resulting in $X_t$, serving as input for the diffusion model. The output of the diffusion model is the predicted noise $\epsilon_{\theta}(X_t,t)$.}\label{overview_method}
\end{figure}

To train a diffusion model for an inpainting task, the ground truth image $x_{GT}$, a mask $m$ masking out some healthy tissue, and the masked image $b = x_{GT} \odot \neg m$ are required, where $\odot$ denotes an element-wise multiplication. We first draw a noisy ground truth image $x_t$ using Eq.~ \ref{eqn3}, and concatenate it with $m$ and $b$. The resulting matrix $X_t:= (x_{t}, b, m)$ and the time step $t$ then serve as input for the network $\epsilon_{\theta}$. The diffusion model therefore learns to denoise an image $x_t$ conditioned on the masked image $b$ and the mask $m$. An overview of the training and sampling procedure can be found in Fig. \ref{overview_method}.
The following list provides details of the different modified methods we evaluate:
\begin{itemize}
\item \textit{DDPM 2D slice-wise}: A baseline method as presented in \cite{durrer2024denoising} with the input being defined as $X_t:= (x_{t}, b, m)$. Here, $x_t$, $b$ and $m$ are 2D slices. During training and sampling, only slices with a non-zero mask are considered. Finally, the samples are stacked to a 3D volume.
\item \textit{DDPM 2D seq-pos}: The above baseline method is extended by conditioning on the previous slice and positional embedding. The input is defined as $X_t:= (x_{t}, b, m, x_{prev})$, with $x_t$, $m$ and $b$ being 2D slices, and $x_{prev}$ being the previous ground truth 2D slice (without noise) during training, or the previously sampled slice during sampling. In addition, we use a position embedding of the slice index. We perform slice-by-slice sampling of non-zero mask slices, where each slice is conditioned on the previous by concatenation, and the samples are stacked to a 3D volume in the end.
\item \textit{DDPM Pseudo3D}: A pseudo-3D method as described in \cite{zhu2023make}, modified for inpainting. Pseudo-3D convolutions result from 2D convolutional layers followed by 1D convolutions in the z-axis. The input is defined as $X_t:= (x_{t}, b, m)$ with $x_t$, $b$ and $m$ being stacks of 2D slices. In contrast to \cite{zhu2023make}, we apply the model in the image space and directly use the pseudo-3D convolutions without the proposed fine-tuning strategy used by \cite{zhu2023make}.
\item \textit{DDPM 3D mem-eff}: A memory efficent 3D diffusion model as presented in \cite{bieder2023memory}. The input is defined as $X_t:= (x_{t}, b, m)$. We decided to use the memory-efficent architecture by \cite{bieder2023memory} for the 3D model in the image space as it allowed using two residual blocks per scale, which was not possible if we simply replaced the 2D convolutions by 3D convolutions in the baseline model. For further implementation details see Table \ref{tab1}.
\item \textit{LDM 3D}: A 3D latent diffusion model as presented in \cite{khader2022medical}. The input is defined as $X_t:= (x_{\text{lat}, t}, b_{\text{lat}}, m_{\text{lat}})$, with $x_{\text{lat}, t}$, $b_{\text{lat}}$, and $m_{\text{lat}}$ being the latent representations of $x_{GT}$, $b$, and $m$. These latent representations are obtained through an autoencoder (AE) following a VQ-GAN implementation. The diffusion model in the latent space is less memory-intense than in the image space. The AE required to obtain the latent representations, however, exceeds the available GPU memory for this experiment (\SI{40}{\giga\byte}) at the initial image resolution. Therefore, downsampling of the input volume $x_{GT}$ was required.
\item \textit{WDM 3D}: A 3D wavelet diffusion model as presented in \cite{friedrich2024wdm}. The input is defined as $X_t:= (x_{\text{wav}, t}, b_{\text{wav}}, m_{\text{wav}})$, with $x_{\text{wav}, t}$, $b_{\text{wav}}$, and $m_{\text{wav}}$ being the concatenated wavelet coefficients of $x_{GT}$, $b$, and $m$. An inverse wavelet transform is applied to reconstruct the images from the predicted $x_{\text{wav}, 0}$.
\end{itemize}

\section{Experiments}
\subsubsection{Datasets}
We train all methods on the publicly available dataset from the \emph{BraTS 2023 Local Synthesis of Healthy Brain Tissue via Inpainting Challenge} \cite{baid2021rsna,bakas2017advancing,bakas2018identifying,karargyris2023federated,kofler2023brain,menze2014multimodal}, hereinafter called "BraTS dataset". The dataset contains T1-weighted brain MR scans showing tumors. The public training data comprises 1251 scans, and the testing data contains 568 scans. Only healthy tissue is cropped out of the T1 scans for training, and the complete T1 scans are used as ground truth. All scans have an initial resolution of $240\times240\times155$ and are cropped and padded in the respective dimensions to fit the required input dimension of the different models. Additionally, due to limitations in the available GPU memory, the input for the AE of the \textit{LDM 3D} has to be downsampled by average pooling to a resolution of $128\times128\times128$. We remove the top and bottom 0.1 percentile of voxel intensities and normalize the voxel values between 0 and 1. For \textit{LDM 3D} and \textit{WDM 3D}, we normalize the voxel values between -1 and 1, following their original implementation.

We further evaluate the best-performing method, \textit{DDPM Pseudo3D}, on an additional brain MR dataset originating from a multicentre MS cohort study \cite{disanto2016swiss,sinnecker2022brain}, hereinafter called "MS dataset". The participants' data is not publicly available due to data privacy protection. Written consent was obtained from all participants, and the data was coded (i.e., pseudonymized) at the time of the patients' enrollment.
We create synthetic MS lesions by applying the lesion masks of 20 MS patients to 20 healthy scans. For this, we skull-strip the images and elastically register randomly chosen pairs of one patient scan and one healthy scan. We then apply the obtained deformation map to the lesion masks and generate masked healthy scans by applying the now registered masks to the healthy control scans. This process provides a ground truth for each of the required lesion fillings. All scans have an initial shape of $192\times240\times256$ and are cropped and zero-padded to a shape of $224\times224\times224$. For all scans, we remove the top and bottom 0.1 percentile of voxel intensities and normalize them between 0 and 1. Of the resulting 20 images containing synthetic lesions, ten are used for fine-tuning of the \textit{DDPM Pseudo3D} and ten for testing. 

\subsubsection{Implementation Details}
\textit{DDPM 3D mem-eff} is trained on an NVIDIA A100 (\SI{80}{\giga\byte}) GPU, the remaining models on an NVIDIA A100 (\SI{40}{\giga\byte}) GPU. Table \ref{tab1} provides implementation details for training the different methods on the BraTS dataset and their sampling speeds. All diffusion models are trained for $3.25\cdot10^{5}$ iterations with $T=1000$ and a linear noise schedule. The AE part of \textit{LDM 3D} is trained for 6.025$\cdot10^{5}$ iterations on input volumes of shape $128^{3}$ with batch size 2, learning rate $3\cdot10^{-4}$, and codebook size 8, requiring \SI{20.97}{\giga\byte} of memory. For the evaluation of the MS dataset, the \textit{DDPM Pseudo3D} is fine-tuned for $10^{5}$ additional iterations on the MS fine-tuning set.
\begin{table}
\centering
\caption{Implementation details of the evaluated methods using the BraTS data set. Res = input resolution, BS = batch size, LR = learning rate, RB = residual blocks per scale, BC = base channels, Mem = GPU memory required for training, Sam = sampling time for one example image. The BS of \textit{DDPM Pseudo3D} is the number of consecutive slices processed simultaneously.}\label{tab1}
\begin{tabular}{p{0.26\textwidth}|>{\centering}p{0.07\textwidth}|>{\centering}p{0.07\textwidth}|>{\centering}p{0.1\textwidth}|>{\centering}p{0.07\textwidth}|>{\centering}p{0.07\textwidth}|>{\centering}p{0.14\textwidth}|c}
Method & Res & BS & LR & RB & BC & Mem [\SI{}{\giga\byte}] & Sam [\SI{}{\min}] \\
\hline
DDPM 2D slice-wise & $224^{2}$ & $16$ & $10^{-4}$ & $2$ & $128$ & $33.97$ & $20$ \\
DDPM 2D seq-pos & $224^{2}$ & $16$ & $10^{-4}$ & $2$ & $128$ & $33.97$ & $25$ \\
DDPM Pseudo3D & $224^{2}$ & $16$ & $10^{-4}$ & $2$ & $128$ & $40.17$ & $25$ \\
DDPM 3D mem-eff & $256^{3}$ & \hphantom{0}$1$ & $10^{-5}$ & $2$ & \hphantom{0}$29$ & $78.84$ & $20$ \\
LDM 3D (DDPM) & $\hphantom{0}32^{3}$ & $16$ & $10^{-4}$ & 4 & \hphantom{0}$32$ & $27.49$ & \hphantom{0}$1$\\
WDM 3D & $256^{3}$ & \hphantom{0}$1$ & $10^{-5}$ & $2$ & \hphantom{0}$64$ & $32.34$ & \hphantom{0}$5$\\
\end{tabular}
\end{table}

\subsubsection{Evaluation Metrics}
Following \cite{kofler2023brain}, we evaluate the performance on the BraTS test set using SSIM, MSE, and PSNR between the healthy ground truth and inpaintings by the different methods. The generated images are post-processed to match the test images' resolution and intensity values. To determine if the top-performing method enhances clinically relevant tasks, we use the MS test set and evaluate the segmentations of cerebrospinal fluid (CSF), grey matter (GM), and white matter (WM) obtained by FMRIB Software Library (FSL) FAST \cite{zhang2001segmentation,jenkinson2012fsl}. The ground truth is provided by the segmentation of  $x_{GT}$ using FSL FAST. We compute the Dice scores comparing the ground truth and the segmentations of images containing the inpainting of \textit{DDPM Pseudo3D}, top-performing on the BraTS test data, or the inpainting by \textit{FSL lesion filling} \cite{battaglini2012evaluating}, respectively. All scores are calculated over the masked regions of the images only. 

\section{Results}
In Table \ref{tab2}, we compare all evaluated methods on the BraTS test set in terms of SSIM, MSE, and PSNR scores. We observe that \textit{DDPM Pseudo3D} outperforms all other methods. In general, the 3D methods perform worse than the 2D and pseudo-3D methods. 
\begin{table}

\caption{Comparison of the evaluated methods on the BraTS test set, considering the masked regions only. All scores are reported as mean $\pm$ standard deviation. The \textit{DDPM Pseudo3D} achieves the best results across all evaluated metrics. }\label{tab2}

\begin{tabular}{l|>{\centering}p{0.26\textwidth}|>{\centering}p{0.26\textwidth}|c}
Method &  SSIM $\uparrow$ & MSE $\downarrow$ & PSNR $\uparrow$ \\
\hline
DDPM 2D slice-wise & 0.7847 $\pm$ 0.1546 & 0.0160 $\pm$ 0.0118 & 18.7131 $\pm$ 3.0824 \\
DDPM 2D seq-pos & 0.7732 $\pm$ 0.2042  & 0.0420 $\pm$ 0.1136 & 18.4529 $\pm$ 5.2799\\
DDPM Pseudo3D & \bftab {0.8527} $\pm$ 0.1196 & \bftab {0.0103} $\pm$ \bftab {0.0107} & \bftab {20.9258} $\pm$ \bftab {3.3835} \\
DDPM 3D mem-eff & 0.7010 $\pm$ 0.1609 & 0.0523 $\pm$ 0.0222 & 14.1384 $\pm$ 3.1795 \\
LDM 3D & 0.5961 $\pm$ 0.1247 & 0.0700 $\pm$ 0.0280 & \hphantom{0}8.7076 $\pm$ 1.8899\\
WDM 3D & 0.6074 $\pm$ 0.1642 & 0.1060 $\pm$ 0.0757 & 10.5693 $\pm$ 3.2045\\
\end{tabular}

\end{table}

\begin{figure}
\centering
\includegraphics[width=\textwidth]{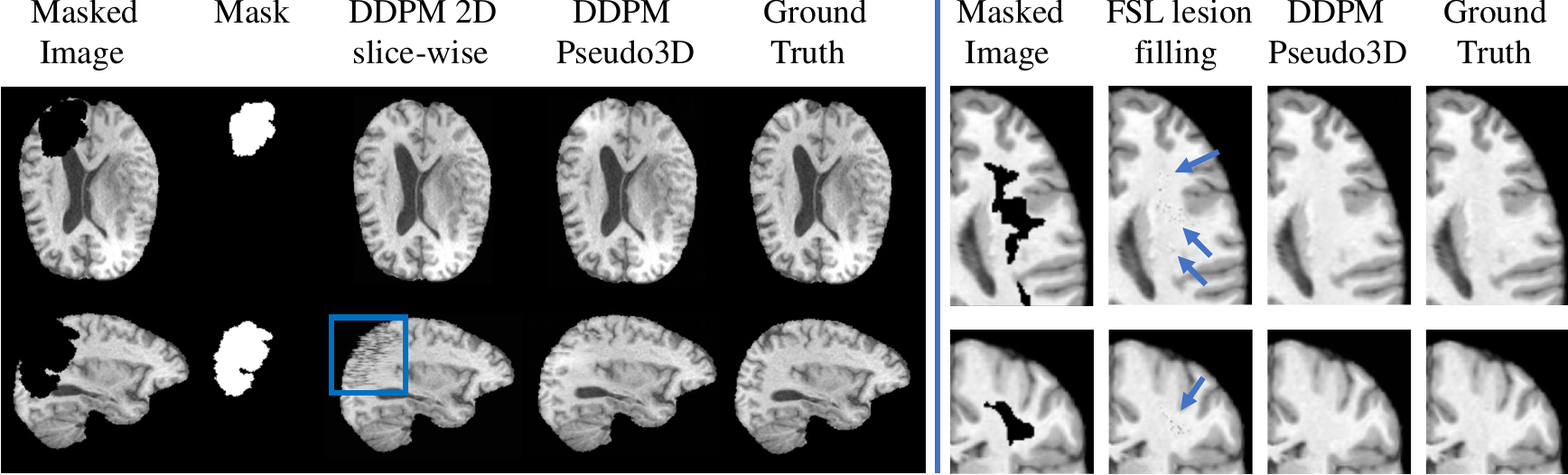}
\caption{Left: Qualitative comparison of \textit{DDPM 2D slice-wise} and \textit{DDPM Pseudo3D} on an example of the BraTS test set. Right: Magnified view of the inpainting by \textit{FSL lesion filling} and \textit{DDPM Pseudo3D} on an example of the MS test set. The blue box and the blue arrows indicate image artifacts. }\label{exemplaryimages}
\end{figure}

For qualitative comparison, we show an example of the inpainting by the baseline \textit{DDPM 2D slice-wise} and the best-performing method \textit{DDPM Pseudo3D} on an image of the BraTS test set in Fig. \ref{exemplaryimages} (left). We observe that \textit{DDPM 2D slice-wise} generates stripe artifacts (indicated by the blue box), while \textit{DDPM Pseudo3D} creates consistent 3D volumes. Examples created by all different methods can be found in the Supplementary Material.

\begin{table}
\centering
\caption{Comparison of the Dice scores for CSF, GM, and WM segmentations in the masked areas for the baseline \textit{FSL lesion filling} and the proposed \textit{DDPM Pseudo3D}, respectively. The ground truth is provided by segmenting the $x_{GT}$ using FSL FAST. All scores are reported as mean $\pm$ standard devation on the MS test set. \textit{DDPM Pseudo3D} outperforms \textit{FSL lesion filling} for all segmentations.}\label{tab3}
\begin{tabular}{p{0.25\textwidth}|>{\centering}p{0.25\textwidth}|>{\centering}p{0.25\textwidth}|c}
Method & CSF & GM & WM \\
\hline
FSL lesion filling & 0.5225 $\pm$ 0.0876 & 0.5853 $\pm$ 0.0877 & 0.9540 $\pm$ 0.0241\\
DDPM Pseudo3D & \bftab {0.8569} $\pm$ \bftab {0.0383} & \bftab {0.8234} $\pm$ \bftab {0.0609} & \bftab {0.9846} $\pm$ \bftab {0.0078}\\
\end{tabular}
\end{table}
 As the \textit{DDPM Pseudo3D} performs best among all methods evaluated on the BraTS test set, we fine-tune it for MS lesion filling to evaluate it for the downstream task of brain tissue segmentation compared to the baseline \textit{FSL lesion filling}. Table \ref{tab3} shows the Dice scores of the inpainting generated by \textit{DDPM Pseudo3D} and \textit{fsl lesion filling} on the MS test set regarding GM, WM, and CSF segmentation.
 In Fig. \ref{exemplaryimages} (right), we present a magnified view of the inpainting by \textit{FSL lesion filling} and \textit{DDPM Pseudo3D} on an image of the MS test set. 

\section{Discussion}
As can be seen in Table \ref{tab2}, \textit{DDPM Pseudo3D} shows the best performance in 3D inpainting. Although \textit{DDPM 3D mem-eff} is trained on a GPU with more memory, it does not show comparable results to the 2D and pseudo-3D methods. In general, all evaluated 3D methods perform worse than the 2D and pseudo-3D methods. We assume that this could be due to the lower number of base channels in the 3D models compared to the 2D and pseudo-3D models. Further research is required to investigate potential limitations regarding model architecture.
Concerning \textit{LDM 3D}, we assume that the performance of the AE is also limited by the GPU memory, leading to comparably low inpainting scores. This assumption is underlined by the poor reconstruction performance when only encoding and decoding an image without performing any diffusion process in between, an example is provided in the Supplementary Material. \textit{WDM 3D} seems to be a fast alternative, requiring fewer resources than common 3D models but still obtaining better scores than, e.g., \textit{LDM 3D}.
As shown in Table~\ref{tab3}, \textit{DDPM Pseudo3D} outperforms \textit{FSL lesion filling} regarding all segmentations. Additionally, the right part of Fig. \ref{exemplaryimages} shows the grainy white matter artifacts generated by \textit{FSL},  while \textit{DDPM Pseudo3D} provides clear and smooth inpaintings. We conclude that the proposed inpainting approach improves clinically relevant downstream tasks compared to the widely used baseline method \textit{FSL}.
 
\section{Conclusion}

 We have modified various state-of-the-art diffusion models to fit the task of 3D healthy tissue inpainting, and conducted a comprehensive comparison between the methods.
Our findings demonstrate that a pseudo-3D approach surpasses the 2D and 3D approaches, and can be efficiently trained on a single \SI{40}{\giga\byte} GPU. 
While 2D methods suffer from limited information about neighboring slices, leading to volume inconsistencies, memory consumption remains a challenge for 3D models operating in the image space.
 The sampling duration is still a general problem, as 
 all comparing methods that operate on the image space have sampling times in the range of 20 minutes per volume.
 Wavelet diffusion models present a promising alternative and require further exploration. Our work offers solutions for consistent 3D healthy tissue inpainting, crucial for clinically relevant downstream tasks such as brain tissue segmentation.

%

\end{document}


\title{Supplementary Material}
%
%

%
\author{}

\institute{}

\maketitle              
%

%


\section{Autoencoder \textit{LDM 3D}}
\begin{figure}[H]
    \centering
    \resizebox{.82\textwidth}{!}{
        \begin{tikzpicture}
            \fill (0.1, 0.1) rectangle (2.125, 3.125);
            \node[] at (0,0) [anchor=south west] {\includegraphics[width=1cm]{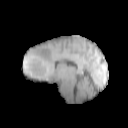}};
            \node[] at (0,1) [anchor=south west] {\includegraphics[width=1cm]{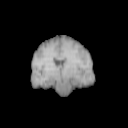}};
            \node[] at (0,2) [anchor=south west] {\includegraphics[width=1cm]{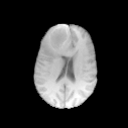}};
            \node[] at (1,0) [anchor=south west] {\includegraphics[width=1cm]{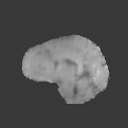}};
            \node[] at (1,1) [anchor=south west] {\includegraphics[width=1cm]{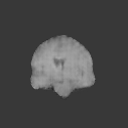}};
            \node[] at (1,2) [anchor=south west] {\includegraphics[width=1cm]{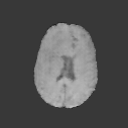}};
            \node[] at (0.6,3.125) [anchor=south]  {\scalebox{.5}{\tiny \vphantom{$\|_{a}^{b}$}Input}};
            \node[] at (1.6,3.125) [anchor=south]  {\scalebox{.5}{\tiny \vphantom{$\|_{a}^{b}$}Reconstruction}};
            
        \end{tikzpicture}
    }
    \caption{Example of an image of the BraTS test set encoded and decoded by the autoencoder used for the \textit{LDM 3D}. The reconstruction shows less detailed structures than the input image.}

\end{figure}

\section{Exemplary Inpainting Results}
\begin{figure}[H]
    \centering
    \resizebox{.84\textwidth}{!}{
        \begin{tikzpicture} 
           \fill (0.1, 0.1) rectangle (3.125, 9.125);

            \node[] at (0.1, 9.1)    [anchor=south west]{\tiny \vphantom{$\|_{a}^{b}$}Axial};
            \node[] at (1.1, 9.1)    [anchor=south west]{\tiny \vphantom{$\|_{a}^{b}$}Sagittal};
            \node[] at (2.1, 9.1)    [anchor=south west]{\tiny \vphantom{$\|_{a}^{b}$}Coronal};

            \node[] at (-2.5, 0.6)  [anchor=west]        {\tiny{Ground Truth}};
            \node[] at (0, 0)       [anchor=south west]  {\includegraphics[width=1cm]{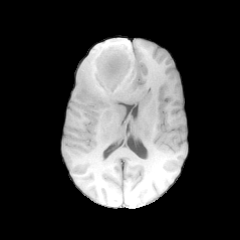}};
            \node[] at (1, 0.2)     [anchor=south west]  {\includegraphics[width=1cm]{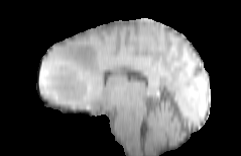}};
            \node[] at (2, 0.2)     [anchor=south west]  {\includegraphics[width=1cm]{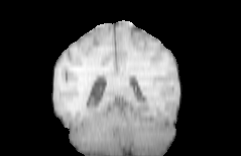}};
            
            \node[] at (-2.5, 1.6)  [anchor=west]        {\tiny{WDM 3D}};
            \node[] at (0, 1)       [anchor=south west]  {\includegraphics[width=1cm]{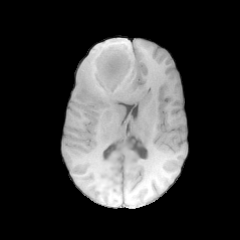}};
            \node[] at (1, 1.2)     [anchor=south west]  {\includegraphics[width=1cm]{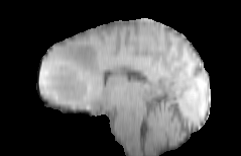}};
            \node[] at (2, 1.2)     [anchor=south west]  {\includegraphics[width=1cm]{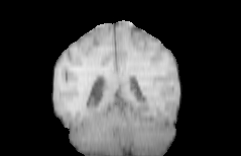}};

            \node[] at (-2.5, 2.6)  [anchor=west]        {\tiny{LDM 3D}};
            \node[] at (0, 2)       [anchor=south west]  {\includegraphics[width=1cm]{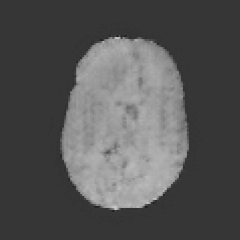}};
            \node[] at (1, 2.2)     [anchor=south west]  {\includegraphics[width=1cm]{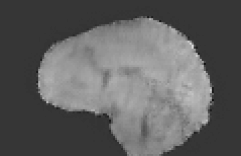}};
            \node[] at (2, 2.2)     [anchor=south west]  {\includegraphics[width=1cm]{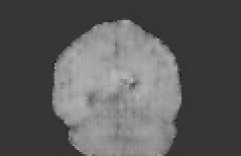}};

            \node[] at (-2.5, 3.6)  [anchor=west]        {\tiny{DDPM 3D mem-eff}};
            \node[] at (0, 3)       [anchor=south west]  {\includegraphics[width=1cm]{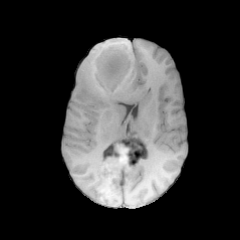}};
            \node[] at (1, 3.2)     [anchor=south west]  {\includegraphics[width=1cm]{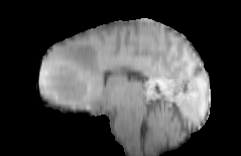}};
            \node[] at (2, 3.2)     [anchor=south west]  {\includegraphics[width=1cm]{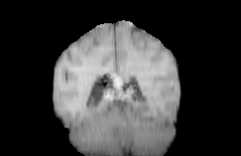}};

            \node[] at (-2.5, 4.6)  [anchor=west]        {\tiny{DDPM Pseudo3D}};
            \node[] at (0, 4)       [anchor=south west]  {\includegraphics[width=1cm]{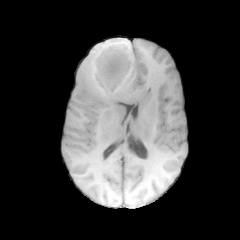}};
            \node[] at (1, 4.2)     [anchor=south west]  {\includegraphics[width=1cm]{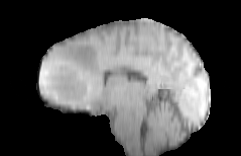}};
            \node[] at (2, 4.2)     [anchor=south west]  {\includegraphics[width=1cm]{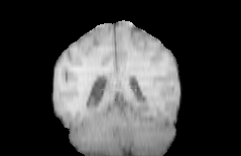}};

            \node[] at (-2.5, 5.6)  [anchor=west]        {\tiny{DDPM 2D seq-pos}};
            \node[] at (0, 5)       [anchor=south west]  {\includegraphics[width=1cm]{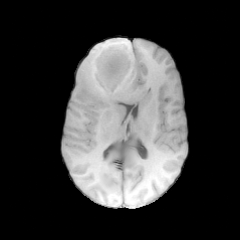}};
            \node[] at (1, 5.2)     [anchor=south west]  {\includegraphics[width=1cm]{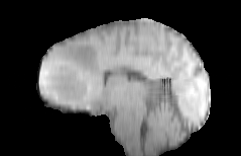}};
            \node[] at (2, 5.2)     [anchor=south west]  {\includegraphics[width=1cm]{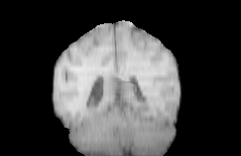}};

            \node[] at (-2.5, 6.6)  [anchor=west]        {\tiny{DDPM 2D slice-wise}};
            \node[] at (0, 6)       [anchor=south west]  {\includegraphics[width=1cm]{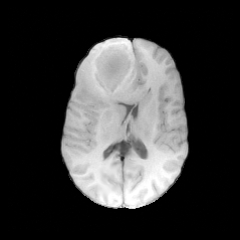}};
            \node[] at (1, 6.2)     [anchor=south west]  {\includegraphics[width=1cm]{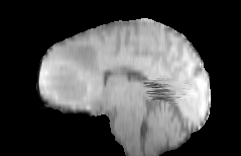}};
            \node[] at (2, 6.2)     [anchor=south west]  {\includegraphics[width=1cm]{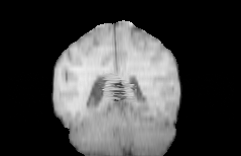}};

            \node[] at (-2.5, 7.6)  [anchor=west]        {\tiny{Mask}};
            \node[] at (0, 7)       [anchor=south west]  {\includegraphics[width=1cm]{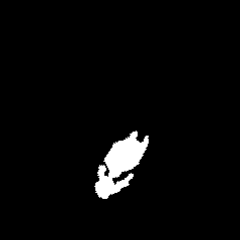}};
            \node[] at (1, 7.2)     [anchor=south west]  {\includegraphics[width=1cm]{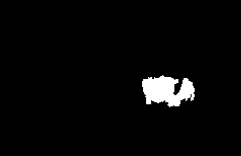}};
            \node[] at (2, 7.2)     [anchor=south west]  {\includegraphics[width=1cm]{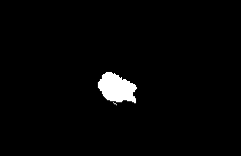}};

            \node[] at (-2.5, 8.6)  [anchor=west]        {\tiny{Masked Image}};           
            \node[] at (0, 8)       [anchor=south west]  {\includegraphics[width=1cm]{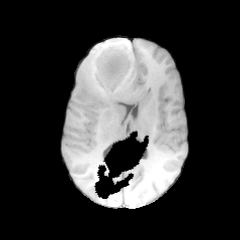}};
            \node[] at (1, 8.2)     [anchor=south west]  {\includegraphics[width=1cm]{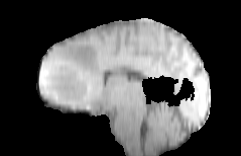}};
            \node[] at (2, 8.2)     [anchor=south west]  {\includegraphics[width=1cm]{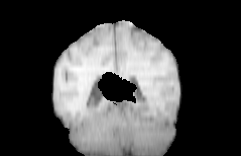}};
    	\end{tikzpicture}
     }
    \caption{Exemplary inpainting results on an image of the BraTS test set generated by all evaluated methods.}
\end{figure}